\documentclass[12pt,aps,prb,preprint]{revtex4-1}   

\usepackage{amsmath}    
\usepackage{amsfonts}
\usepackage{mathrsfs, mathtools, dsfont, tensor}

\newcommand{\Lag}{\mathcal{L}}

\DeclarePairedDelimiter{\bra}{\langle}{\rvert}
\DeclarePairedDelimiter{\ket}{\lvert}{\rangle}

\DeclarePairedDelimiter{\abs}{\lvert}{\rvert}

\DeclareMathOperator{\diag}{diag}

\newcommand{\braket}[2]{\langle#1|#2\rangle}

\begin{document}


\title{A proof of Schr\"odinger's equation}
\author{Howard Covington}
 \affiliation{Isaac Newton Institute for Mathematical Sciences, Cambridge, England}
 \email{hcovington@newton.ac.uk}   
\date{\today}

\begin{abstract}
Schr\"odinger's equation for a single particle is proved from the assumption that dynamics can be formulated in a space whose curvature is the electromagnetic force.
\end{abstract}

\maketitle

Einstein's general theory of relativity brought to physics the idea that the gravitational force is geometric in origin and may be identified with the curvature of spacetime. The mathematics needed to apply more generally the idea that force can be identified with curvature is the theory of vector bundles. This is a part of a theory that was completed by mathematicians in the 1950s. By that time, physicists had already discovered its main ideas for themselves as gauge theory \cite{O'Raifeartaigh:1997ia}. It is nonetheless worthwhile to seek to express the dynamics of particle in a space whose curvature can be identified with the electromagnetic force, since this leads to a proof of Schr\"odinger's equation and so to elementary quantum mechanics.

General relativity describes curvature by attaching to each point in spacetime a vector space called the tangent space at that point. Differential relationships between vectors in neighbouring tangent spaces are defined by the Christoffel symbols $\Gamma\indices{ _\mu^\alpha_\beta}(x)$. A covariant derivative $D_\mu$  acts on a vector field $v^\alpha(x)$ according to
\begin{equation}
  D_\mu v^\alpha = \partial_\mu v^\alpha + \Gamma\indices{_\mu^\alpha_\beta} v^\beta.
\end{equation}
The covariant derivative may be used to define parallel transport of a vector. Curvature is the change in a vector when it is parallel-transported around an infinitesimal closed loop. In the theory of vector bundles \cite{Baez:1995sj} a space $S$ is created by attaching to each point in spacetime a vector space that can carry the representation of a group. Differential relationships between vectors in neighbouring vector spaces are defined by a connection $A\indices{_\mu^a_b}(x)$. If the group is the group of rotations around a unit circle, then the vector space is equivalent to the space of complex numbers. The simplest object that can inhabit $S$ in this case is a complex-valued function $\psi(x)$. The connection now becomes wholly imaginary and the covariant derivative $D_\mu$  acts on  $\psi$ according to
\begin{equation}\label{eq:D_mu_psi}
  D_\mu\psi = \partial_\mu \psi + i\gamma A_\mu \psi,
\end{equation}
where $A_\mu$  is real and $\gamma$ is a real number introduced for convenience. The curvature, $F_{\mu\nu}$, of $S$ is given by $F_{\mu\nu} = \partial_\mu A_\nu - \partial_\nu A_\mu$. It is invariant under group operations, which in this context are called gauge transformations, and satisfies the second Maxwell equation  $\epsilon^{\alpha\beta\mu\nu}\partial_\beta F_{\mu\nu}=0$.  This establishes that the Faraday tensor is the curvature of a space $S$ whose vector space is the space of complex numbers.

The question is now how to define the dynamics of a test particle in this space. Parallel transport in $S$ is defined in relation to the vector space of complex numbers rather than in relation to the tangent space of a curve so, in contrast to general relativity, $A_\mu$  cannot be used to define a class of curves that parallel transport their tangent vectors. So geodesics are not defined by $A_\mu$ and there is no natural equation for a test particle trajectory. We therefore have to look elsewhere for a formulation of dynamics. To this end we seek a class of functions $\psi$ that have the dynamical properties of classical particles.

Suppose $\psi$ is given by a differential equation. $A_\mu$  defines differential relations, so it will appear in this equation. $\psi$ will therefore appear to interact with an electromagnetic field, so it must have a charge and will be a source for $A_\mu$. The evolution of $\psi$ and $A_\mu$  will therefore be coupled together. Since the electromagnetic field  may be specified by a least action problem \cite{Saletan:1998:569}, $\psi$ and $A_\mu$  can be coupled together consistently if $\psi$ is also specified by a least action problem. The problem of defining $\psi$ is therefore equivalent to defining an appropriate Lagrangian for $\psi$. The formalism of vector bundles now comes into its own, because the Lagrangian for $\psi$ in the presence of the connection $A_\mu$  can be obtained from the free Lagrangian $\Lag$ for the free field $\psi$ when the connection is zero by the substitution of $\partial_\mu$  by $D_\mu$  given by \eqref{eq:D_mu_psi}. This is just the minimal coupling prescription from quantum field theory appearing here in geometric form.

Since $\psi$ has charge it must represent charged matter. Accordingly, it should be possible to construct a specific $\psi$ that represents a classical charged particle. $\Lag$ will then determine a function that represents a free particle. The dynamics of a free particle are specified by its 4-momentum, $P^\mu$, its angular momentum and its position. Classically, the 4-momentum is constant, unique, has a positive energy and is otherwise constrained only by the requirement from the special theory of relativity that $P^2 = m^2c^2$. We therefore seek a free Lagrangian that enables $\psi$ to incorporate these properties, insofar as it is possible to do so.

To make progress we use Noether's theorem \cite{Saletan:1998:565}. Subject to appropriate boundary conditions, this theorem associates with each symmetry of a least action problem a conserved current $j^\mu(x)$ satisfying $\partial_\mu j^\mu=0$ and a constant of the dynamics equal to $\int j^0d^3x$. When the free electromagnetic field is specified by a least action problem, the four constants defined by Noether's theorem that arise from invariance under spacetime translations are equal to the components of its 4-momentum \cite{Rindler:2001}. Similarly, the six constants that arise from invariance under rotations and Lorentz transformations are the components of the field's angular momentum tensor.

The electromagnetic field and the function $\psi$ are going to be coupled together in a single least action problem, so for consistency, $\psi$'s momentum and angular momentum must be defined by applying Noether's theorem to the least action problem for $\psi$. For convenience we refer to $\psi$'s momentum defined in this way as its Noether momentum.

We now have enough substantially to determine $\Lag$. $\Lag$ is a functional of $\psi$ and its derivatives. For the least action problem to be well defined, $\Lag$ must be real. For $\psi$ to have dynamical behaviour $\Lag$ must contain at least $\psi$ and its first derivatives. So that the least action problem is formulated covariantly and is independent of gauge transformations, $\Lag$ must be a Lorentz- and gauge-invariant scalar. For $\Lag$ to be gauge-invariant each term in $\Lag$ must be at least a covariant bilinear combination of  $\psi^*$ (or a derivative of  $\psi^*$) and  $\psi$ (or a derivative of $\psi$ ) or a higher order gauge-invariant combination. So that $\Lag$ captures the dynamics of a free particle, it must include the particle's mass $m$.

The simplest Lagrangian that satisfies these requirements consists of covariant bilinear combinations of a scalar function $\psi$ and its complex conjugate $\psi^*$ and their derivatives up to first order. It therefore contains two arbitrary real constants, denoted for convenience $\hbar$ and $\mu^2$, and may be expressed as
\begin{equation}
    \Lag = \hbar\left(\partial_\mu\psi^*\partial^\mu\psi - \mu^2\psi^*\psi\right).
\end{equation}
The Euler-Lagrange equation for the least action problem is the Klein-Gordon equation $(\partial^2 + \mu^2)\psi = 0$. The general solution contains terms that would have both positive and negative energy \cite{Maggiore:1998:51}. To avoid the negative energy solutions we choose to consider only solutions of the form
\begin{equation}\label{eq:psi_positive}
    \psi = \int \frac{d^3k}{\sqrt{(2\pi)^3 2k^0}}a(\vec{k})e^{-ikx}.
\end{equation}
When we examine how $\psi$ interacts we will want to expand it as a sum over plane wave states.  So for analysing interactions, this choice limits the theory we are constructing to the low energy approximation where the functions $e^{i\vec{k}\cdot\vec{x}}$ are complete on a surface of constant time.

The value for $\gamma$ in \eqref{eq:D_mu_psi} can now be fixed in terms of $\psi$'s charge. Since $\Lag$ is gauge invariant, it is invariant under the phase symmetry $\psi\to e^{-i\lambda}\psi$, where $\lambda$ is a real number. By Noether's theorem this phase symmetry gives rise to a vector $j^\mu$ satisfying $\partial_\mu j^\mu = 0$ where\cite{Saletan:1998:577}
\begin{equation}\label{eq:j_mu_def}
  j^\mu = i\left(\psi^* \frac{\partial \Lag}{\partial(\partial_\mu\psi^*)} -
                    \frac{\partial \Lag}{\partial(\partial_\mu\psi)}\psi\right).
\end{equation}
In the presence of a connection, $\partial_\mu$  is replaced by $D_\mu$  in $\Lag$. Denoting the resulting Lagrangian by $\Lag_D$, it follows from \eqref{eq:D_mu_psi} that
\begin{equation}\label{eq:j_mu}
    \frac{\partial\Lag_D}{\partial A_\mu}\biggr\rvert_{A=0} =
    - \gamma i\left(\psi^* \frac{\partial \Lag}{\partial(\partial_\mu\psi^*)} -
                    \frac{\partial \Lag}{\partial(\partial_\mu\psi)}\psi\right) = -\gamma j^\mu.
\end{equation}
In the presence of an electromagnetic field, the complete Lagrangian for the least action problem is the sum of the electromagnetic Lagrangian, $-\tfrac{1}{4}F_{\mu\nu}F^{\mu\nu}$, and $\Lag_D$. When $A_\mu$ is varied to determine the least action, the electromagnetic Lagrangian will contribute $\partial_\nu F^{\nu\mu}$ and $\Lag_D$ will contribute $\tfrac{\partial L_D}{\partial A_\mu}$. Inserting appropriate constants\footnote{The convention used is $g^{\mu\nu} = \diag(1, -1, -1, -1)$ and SI units for the electromagnetic field so that the action is $\int \Lag_{\text{EM}}d^3xdt=c^{-1}\int\Lag_{\text{EM}}d^4x$ where $L_{\text{EM}} = -\tfrac{1}{4\mu_0}F_{\mu\nu}F^{\mu\nu} + J^\mu A_\mu $ and $J^\mu =(c\rho ,\vec{j})$. }, the condition for the action to be minimized for variations in $A_\mu$ will be
\begin{equation}\label{eq:first_maxwell}
  \frac{1}{\mu_0} \partial_\nu F^{\nu\mu} = -c \frac{\partial \Lag_D}{\partial A_\mu}.
\end{equation}
By the first Maxwell equation, $\partial_\nu F^{\nu\mu} = \mu_0 J^\mu$, the right hand side of \eqref{eq:first_maxwell} must be identified with the electromagnetic current density associated with $\psi$ and the zeroth component with $\psi$'s charge density. So from \eqref{eq:j_mu}, $c\gamma j^\mu$  can be identified with the electromagnetic current density for the free field $\psi$. Integrating $c\gamma j^0$ establishes that the charge $q$ of the free field $\psi$  satisfies $ q =\gamma \int j^0 d^3x$.

The problem with this is that $\int j^0 d^3x$ is different for each function $\psi$ given by the same $\Lag$, so $\gamma$ is not constant and the covariant derivative is not properly defined. We can solve this problem in the following way. Using the expression \eqref{eq:psi_positive} for $\psi$ in \eqref{eq:j_mu} we find that $\int  j^0 d^3x = \hbar \int d^3k \abs{a(\vec{k})}^2$. Since the Klein-Gordon equation is linear, if $\psi$ is a solution then so is $\sigma \psi$, where $\sigma$ is any complex number, and for this solution $\int j^0 d^3x$ is multiplied by $\abs{\sigma}^2$. We now agree to use a particular value of $\sigma$ to represent all $\psi$s that are generated by varying $\sigma$. By choosing this value to be $\sigma  = \left( \int d^3k \abs{a(\vec{k})}^2\right)^{-1/2}$, we arrange that  $\int j^0 d^3x=\hbar$, ensuring that the covariant derivative is independent of $\psi$. As a consequence we have $\gamma = \tfrac{q}{\hbar}$ and the covariant derivative defined in \eqref{eq:D_mu_psi} becomes
\begin{equation}\label{eq:em_cov_der}
  D_\mu = \partial_\mu + i \frac{q}{\hbar} A_\mu.
\end{equation}
Defining normalized coefficients $\alpha(\vec{k})$ by $\alpha(\vec{k}) \equiv \sigma a(\vec{k})$, $\psi$ takes the standard form
\begin{equation}\label{eq:psi_standard}
  \psi = \int \frac{d^3k}{\sqrt{(2\pi)^3 2k^0}} \alpha(\vec{k}) e^{-ikx}.
\end{equation}
Having determined $\gamma$, we now turn to $\psi$'s momentum. From the translation invariance of the least action problem, $\psi$'s Noether momentum, $P^\mu$, is given \cite{Saletan:1998:567} by
\begin{equation}\label{eq:noether_momentum}
  P^\mu = \int d^3x \left( \frac{\partial\Lag}{\partial(\partial_0 \psi)} \partial^\mu \psi +
                        \frac{\partial\Lag}{\partial(\partial_0 \psi^*)} \partial^\mu \psi^* -
                        g^{0\mu}\Lag\right).
\end{equation}
To express $P^\mu$  more compactly we use \eqref{eq:j_mu_def} and the fact that $\Lag$ is bilinear to introduce a symbol $\bra{\psi} O \ket{\psi}$, where $O$ is a linear operator, defined by
\begin{equation}\label{eq:braket_operator}
  \bra{\psi} O \ket{\psi} \equiv \frac{i}{\hbar} \int d^3x \left[ \psi^* O
        \left(\frac{\partial\Lag}{\partial(\partial_0 \psi^*)}\right) -
         \frac{\partial\Lag}{\partial(\partial_0 \psi^)} O(\psi) \right]
\end{equation}
We denote $\bra{\psi} I \ket{\psi}$ by $\braket{\psi}{\psi}$, where $I$ is the identity operator, so that $\braket{\psi}{\psi} = 1$. It then follows \cite{Maggiore:1998:51-53} that
\begin{equation}\label{eq:p_mu}
  P^\mu = \bra{\psi} i\hbar \partial^\mu \ket{\psi}.
\end{equation}
With $\psi$ given by \eqref{eq:psi_standard}, then using \eqref{eq:noether_momentum} or \eqref{eq:p_mu} we have
\begin{equation}\label{eq:p_mu_alpha}
  P^\mu = \hbar \int d^3k \abs{\alpha(\vec{k})}^2 k^\mu.
\end{equation}
$P^0$ is positive provided $\hbar$ and $k^0$ are both positive. If $\Lag$ is replaced by $\rho\Lag$, where  $\rho$ is any real number, then the Klein-Gordon equation, and therefore $\psi$, are unchanged but $P^\mu$  is multiplied by $\rho$. If  $\psi$ is to represent a free particle then its Noether momentum must be unique. We must therefore eliminate the freedom to vary $\rho$ in this way. To arrange this we choose a specific $\Lag$ to represent the different $\Lag$'s that are generated by varying $\rho$ by agreeing a unique positive value (as yet unknown) for $\hbar$. If $\psi$ is a plane wave of the form $\psi = \alpha(\vec{k})e^{-ikx}$ then its momentum is just $P^\mu  = \hbar k^\mu$, which is the de Broglie relation.

Just as Noether momentum is defined in relationship to the translational invariance of the least action problem, we can define Noether angular momentum from its rotational invariance. Using the symbol $\bra{\psi} O \ket{\psi} $ defined in \eqref{eq:braket_operator}, the result is a 3-vector $\vec{J}$ of constants \cite{Maggiore:1998:51-53}, where $\vec{J} = \bra{\psi} \vec{x} \times (-i \hbar \vec{\nabla} ) \ket{\psi}$. If we assume that $\psi$ is localized, then it is also meaningful to define a measure $\langle \vec{x} \rangle$ of the position of $\psi$ by $\langle \vec{x} \rangle \equiv \bra{\psi} \vec{x} \ket{\psi}$.

We now turn to the question of how to represent a particle in the theory. If $\psi$ is a plane wave, the momentum condition $P^2 = m^2c^2$  can be satisfied with $P^\mu$  given by $P^\mu  = \hbar k^\mu$  provided that $k^\mu$  satisfies $\hbar^2 k^2= m^2c^2$. This can be arranged by choosing the parameter  $\mu$ in the Klein-Gordon equation to be equal to $mc/ \hbar$. If $\psi$  is the linear combination \eqref{eq:psi_standard}, the momentum condition will not then be satisfied exactly  but will still hold approximately provided the coefficients $\alpha(\vec{k})$ are only non-zero in a narrow range $\delta\vec{k}$ of values for $\vec{k}$ where $\abs{\delta\vec{k}} \ll \abs{\vec{k}}$. Provided that $\hbar$ is sufficiently small, $\psi$ can, in principle, be constructed so that it is both localized on a small scale and its momentum satisfies the momentum condition to a good approximation.

In this way we are led to the proposal that a localized function $\psi$, or wave packet, is a possible representation of a free particle in this theory. From the point of view of the classical dynamics  we started from, this is not an ideal solution since it does not satisfy exactly the requirements we set. But it is the best we can do. For the moment we assume that $\psi$ is always localized on a scale much smaller than the laboratory scale for making observations of particles so that we cannot observe the internal structure of a wave packet.

To ensure that the momentum condition is satisfied we have relied on the form of the Klein-Gordon equation. If higher order derivatives had been included in $\Lag$ this would not have been achieved or, at least, not in a simple way.  If $\Lag$ contained terms that were higher order than bilinear in $\psi^*$ and $\psi$ then the field equations would not be linear. The Noether momentum would then not be proportional to an unconstrained wave vector $k^\mu$. For these reasons we conclude that $\Lag$ must be bilinear in $\psi^*$ and $\psi$   and contain derivatives of first, but no higher, order. A scalar function $\psi$ is only the simplest object that can inhabit $S$. A complex function with several components could equally well be defined on $S$. So that a $\psi$ with several components may be combined into a Lorentz-invariant Lagrangian the components must be drawn from a particular representation of the Lorentz group. There will therefore be a class of relevant Lagrangians, whose members correspond to different irreducible representation, all of which are bilinear and contain only first order derivatives.

For each Lagrangian in this class the arguments leading through to equation \eqref{eq:p_mu_alpha} will be unchanged in character save that $\psi$, $a(\vec{k})$ and $\alpha(\vec{k})$ acquire an index, $I$, that labels components and the form of the Lagrangian will be different for each representation. Since the constant $\int j^0 d^3x$ and the Noether momentum are constructed covariantly, they will both depend on the amplitude $a_I(\vec{k})$ only through the combination $\abs{a(\vec{k})}^2 \equiv \Sigma_I a^*_I(\vec{k})a_I(\vec{k})$, so the reasoning leading to \eqref{eq:em_cov_der} will be the same. Just as $\hbar$ has to be independent of $\psi$ for the covariant derivative to be properly defined, so $\hbar$ must also be independent of the representation from which $\psi$ is drawn and we must always have $\gamma = \tfrac{q}{\hbar}$ in the covariant derivative. Lagrangians for different representations must include appropriate numerical factors to give this result. $\hbar$ will then be the same for all representations and becomes a structural constant of the theory. We consider later the significance of $\psi$ having several components.

We now have everything we need to prove Schr\"odinger's equation. In the presence of an electromagnetic field, the Lagrangian becomes $\Lag_D = \hbar\left[(D_\mu \psi)^*(D^\mu \psi) -\left(\frac{mc}{\hbar}\right)^2 \psi^* \psi\right]$. The Euler-Lagrange equation is $\left(D_\mu D^\mu  + \left(\frac{mc}{\hbar}\right)^2\right)\psi  = 0$. The low-energy approximation to this equation may be found in a standard way \cite{Schiff:2001:469}. Setting $A_0 = \phi/c$ and defining the differential operator $\vec{\pi}$ by $\vec{\pi} \equiv -i\hbar \vec{\nabla}$, the result is Schr\"odinger's equation
\begin{equation}\label{eq:schrodinger}
  i\hbar \frac{\partial \psi}{\partial t} = \left[\frac{1}{2m} (\vec{\pi} - q \vec{A})^2 + q\phi \right] \psi.
\end{equation}
The differential operator on the right hand side of this equation takes the form $H(\vec{\pi},\vec{x})$ where $H(\vec{p},\vec{x})$ is the classical Hamiltonian for a particle in an electromagnetic field \cite{Saletan:1998:204}. We thus establish from first principles that there is a strikingly simple map from classical mechanics to the theory of functions $\psi$ in the low-energy approximation.

The simplicity of this map is no coincidence. The form \eqref{eq:schrodinger} of Schr\"odinger's equation is a direct consequence of defining $\psi$'s momentum by Noether's theorem, of the requirement for the free Lagrangian to be bilinear if $\psi$'s properties are to be like those of a free particle and of the minimal coupling of $\psi$ to the electromagnetic field that results from treating the electromagnetic force as the curvature of a space. That the classical Hamiltonian has the same form follows from a theorem in classical mechanics \cite{Hughes:1992,Castro:1993}, that the Hamiltonian for a broad class of forces that depend on position and/or velocity but not acceleration can be put in the minimal coupling form $H(\vec{p},\vec{x})$. A simple map between classical and quantum mechanics is therefore built in to the construction of both theories. The classical result was discussed in comments on Feynman's proof of Maxwell's equations\cite{Dyson:1990}.

Having found Schr\"odinger's equation we can work back to a free Lagrangian for it and then calculate the symbol $\bra{\psi}O\ket{\psi}$ and thus $\psi$'s Noether momentum. We find that $\bra{\psi}O\ket{\psi}$ is just $\int d^3x \psi^* O \psi$. $\braket{\psi}{\psi}$ can now be used to normalize any $\psi$ and to define a scalar product for any two functions. So, in the low energy approximation all square-integrable functions $\psi$ in the standard form \eqref{eq:psi_standard} are normalized functions, or rays, in Hilbert space \cite{Hassani:2006} and can be treated by the theory of functions and linear operators on such a space. $\psi$ has Noether 3-momentum $\bra{\psi}\vec{\pi}\ket{\psi}$ and angular momentum $\bra{\psi}\vec{x}\times\vec{\pi}\ket{\psi}$, where these are now the familiar quantum mechanical expressions. Its position is $\bra{\psi}\vec{x}\ket{\psi}$. The position and momentum operators satisfy $\left[x^i, \pi^j\right] = i\hbar \delta^{ij}$. By seeking the dynamics of a particle in a space for which the electromagnetic field is its curvature we are therefore led to create a representation of a particle by a ray in Hilbert space and to infer the usual commutation relations for the position and momentum operators.

To complete the proof we consider how $\psi$'s properties might be observed. When $\psi$ interacts with an electromagnetic field it can transfer momentum to the field. Since momentum is conserved, changes in $\psi$'s momentum can in principle be observed by measuring changes to the momentum of the electromagnetic field.  The same remarks apply to $\psi$'s angular momentum. Since $\psi$ is assumed to be localized,  $\bra{\psi}\vec{x}\ket{\psi}$ will be the approximate location of the charge $q$. This can also be observed through the electromagnetic field it gives rise to. So we expect momentum, angular momentum and position to be observables and to be given by the expression  $\bra{\psi}O\ket{\psi}$ where $O$ is the relevant self-adjoint operator in Hilbert space.

A calculation of the scattering of an incoming wave packet from a potential \cite{Schiff:2001} shows, however, that we cannot always assume that $\psi$ is localized. The outgoing wave is a shell whose radius can, in principle, grow indefinitely and so assume laboratory dimensions, making the internal structure of a wave packet available for observation.

Such an observation might be made by seeking to detect $\psi$'s charge $q$ using the sort of equipment whose use led to the definition of charge and the formulation of the laws of electromagnetism. Suppose such a detector has a cavity of dimensions $\delta \vec{y}$ for capturing charge and that $\psi$ is localized within some region $\delta\vec{x}$. Provided  $\delta\vec{y}$ is large enough entirely to contain $\delta\vec{x}$, we may assume that the detector unambiguously registers the charge $q$ and thus registers the detection of $\psi$. For convenience, we may also assume that using the same equipment we can arrange to measure $\psi$'s momentum and its mass.

By the early 1900s it was established that the charge associated with matter came in discrete units of a minimum charge $\pm e$ and that at a small scale there were particles, such as the electron, that carried just this charge and had a characteristic rest mass. Suppose $\psi$ represents one of these particles and is scattered so that it becomes an extended wave front whose breadth is much greater than the width $\delta\vec{y}$ of a detector's cavity. Furthermore, suppose that detectors can be arranged, and the wave front is sufficiently extensive, that it encounters several detectors simultaneously. Since charge does not exist in an amount less than $e$, a particular detector must either register the whole of the charge $e$ or no charge at all. If a detector registers the charge $e$ it will register the detection of $\psi$; otherwise it will register nothing. Moreover, if one detector registers the charge $e$, then since charge is conserved, no other detector can register it. So even if the function $\psi$ simultaneously encounters a number of detectors, at most one detector can register a detection.  But all detectors will potentially respond to $\psi$'s charge in the same way. So to produce the correct result from one detector and to avoid multiple detections, $\psi$  must collapse instantaneously into a new normalized function that is only non-zero within the dimensions $\delta\vec{y}$ of the detector that registers $\psi$'s charge. The detector will then register the charge of this new function.

This conclusion is clearly the consequence of representing a particle that has a property that is both conserved and discrete by a spatially extensive normalized function. We have framed the above argument in terms of charge, but it could equally well have been framed in terms of rest mass or any other observable property we might subsequently discover (such as spin) that is both discrete and conserved.

To see what happens to $\psi$'s momentum when it is detected and collapses, consider $\psi$ in the form of a shell moving outwards from the origin. Such a $\psi$ can be expressed as the standard linear combination \eqref{eq:psi_standard}. Suppose that we cover a sphere centred at the origin with $N$ contiguous charge and momentum detectors numbered from $n = 1, \dots, N$.  Let the $n$th detector subtend a small solid angle $\Delta\Omega(n)$ at the origin. The only components of the combination that will reach the $n$th detector are those whose momentum vectors $\vec{k}$ lie within $\Delta\Omega(n)$. The amount of $\psi$'s Noether momentum, $P^\mu(n)$, attributable to these components is, from \eqref{eq:p_mu_alpha}
\begin{equation}\label{eq:p_mu_n}
  P^\mu(n) = \int_{\vec{k}\in\Delta\Omega(n)} d^3k \abs{\alpha(\vec{k})}^2 \hbar k^\mu
            \approx \abs{\alpha(n)}^2 \hbar k^\mu(n)
\end{equation}
where $\abs{\alpha(n)}^2 \equiv \int_{\vec{k}\in\Delta\Omega(n)} d^3k \abs{\alpha(\vec{k})}^2$ and it has been assumed that $\Delta\Omega(n)$ is so small that $k^\mu$  is the same for all $\vec{k}$ lying within $\Delta\Omega(n)$ and is denoted $k^\mu(n)$.  If the $n$th detector registers a detection, the momentum it will measure will correspond to the new normalized function into which $\psi$ collapses and will therefore be just $\hbar k^\mu(n)$. It is clear from \eqref{eq:p_mu_n} that $\psi$'s Noether momentum $P^\mu$ is the sum of all the $P^\mu(n)$.

Suppose that the momentum $\hbar k^\mu (n)$ is transferred to the particle detector as part of the detection process. Then the balance of the Noether momentum of the scattered particle $\Delta P^\mu(n) = P^\mu - \hbar k^\mu(n)$ must be transferred to the source of the scattering potential so that momentum is conserved. Now suppose this experiment is repeated many times. The rate of transfer of momentum to the source will in principle be observable as a force on the source that is additional to the Lorentz force that arises from the scattering itself. But the laws of electromagnetism do not include such a `detection force', so there is a potential inconsistency. We can reconcile the existence of this force with the failure to detect it in the following way. The experiments that underpin the laws of electromagnetism typically involve large numbers of charged particles. The fields, currents and charges in these laws are averages over these particles. These particles will scatter off one another, so that charge detectors may be presented with the kinds of linear combination we have been considering. The detection force would, however, escape detection if it averaged out to zero over the large number of particles involved in the experiment or over many repetitions of the same experiment.

For $\Delta P^\mu(n)$ to average out to zero the relative frequency of detection by the $n$th detector would have to be $\abs{\alpha(n)}^2$. Since, in principle, particle detectors can be made as small as we like, we may conclude from this argument that the relative frequency of detection of the component $\alpha(\vec{k})e^{-ikx}$ in $\psi$ must be $\abs{\alpha(\vec{k})}^2$.

Although we have examined a  $\psi$ created by scattering, there is nothing special about it other than our ability to detect individual components because they have been separated in space. We infer that detection applied to any linear combination of plane waves, where it is physically possible to detect individual components, operates according to a probabilistic model, where the probability of detection, $\Pr(\vec{k})$, is $\abs{\alpha(\vec{k})}^2$ in order to produce the relative frequency of detection required for consistency.

According to this probability model, the Noether momentum of a linear combination of plane waves must now be interpreted as the average of the observations over many repetitions of the same experiment. We therefore have from \eqref{eq:p_mu} and \eqref{eq:p_mu_alpha}
\begin{equation}
  P^\mu = \bra{\psi} i \hbar \partial^\mu \ket{\psi} =
         \int d^3k \abs{\alpha(\vec{k})}^2 \hbar k^\mu =
         \int d^3k \Pr(\vec{k}) \hbar k^\mu.
\end{equation}
$\bra{\psi} i \hbar \partial^\mu \ket{\psi}$ is thus a statistical expectation value for the observations of the eigenvalues of the operator $i\hbar\partial^\mu$  .

Using the machinery of function theory in the low energy approximation, this argument can be extended to the eigenvalues of other self-adjoint operators in situations where the corresponding eigenfunctions are physically separated and so can be individually detected. In this way we obtain the general form of the quantum mechanical measurement postulate.

Finally, we consider the consequences of starting from a Lagrangian for a $\psi$ that has multiple components. We have already noted that such a Lagrangian will have to be bilinear, so observables will still be calculated from bilinear expressions $\bra{\psi} O \ket{\psi}$. This bilinearity means that the symmetry group of the theory is the covering group of the Lorentz group rather than the Lorentz group itself. The significance of this is that $\psi$  may belong to spinor as well as tensor representations.

The new consequence of replacing the scalar function $\psi$ by a many-component function is that a term of the form $\bra{\psi} i\hbar\vec{S}\ket{\psi}$  is added to the expression $\bra{\psi}\vec{x} \times ( -i\hbar\vec{\nabla})\ket{\psi}$ for the Noether angular momentum, where $\vec{S}$ is a vector of spin matrices. Schr\"odinger's equation is the low energy approximate equation for the many-component  $\psi$ but now includes a corresponding spin term \footnote{This statement can be verified case by case and is easily done so for $\psi$ with up to four components, when the only relevant equations for $\psi$  are the Klein-Gordon, Dirac and Proca equations. I am not aware of a general proof for an arbitrary number of components}. These spin terms can be calculated by working through possible Lagrangians for $\psi$s with different numbers of components. In particular, by using a 4-component spinor, the Dirac equation is obtained and approximated for low energies by Schr\"odinger's equation for a spin-$\tfrac{1}{2}$ particle \cite{Maggiore:1998:73}.

\section{Commentary}
The main features of the elementary quantum mechanics of a single charged particle in an external electromagnetic field, including its spin, can evidently be derived by seeking to construct a space for which the electromagnetic force acts as a curvature and that admits, as far as possible, the dynamics of a free classical particle expressed through a function $\psi$. From this point of view, Schr\"odinger's equation can be interpreted as the low-energy approximate equation for a function with particle-like properties on a space whose curvature is the electromagnetic field. The probabilistic interpretation of the wave function is required to reconcile the existence of discrete, conserved, observable particle properties with the representation of a particle by a spatially extended function. The structure of the resulting theory is in essence what is summarized by the postulates of quantum mechanics.

To arrive at this point of view we needed to assume the special theory of relativity, Maxwell's equations, the uniqueness of momentum and the existence for dynamics of a least action problem. The structure of wave mechanics then follows largely from the intersection of three ideas. The first is an extension to electromagnetism of Einstein's insight that there is a geometric formulation of the gravitational force as curvature. The second is that conservation laws arise from symmetries through Noether's theorem and the third is that particles have some discrete, conserved, observable properties. The representation of a particle by a ray in Hilbert space and the appearance of a constant $\hbar$ in the theory can be traced to the need for the covariant derivative to be properly defined and for the momentum of a free particle to be unique.

Since we have not needed any explicit quantum mechanical assumptions, it may appear that this proof obtains Schr\"odinger's equation from classical theory. This appearance is misleading and comes about because, once we have constructed a space whose curvature is the electromagnetic force, we seek to formulate particle dynamics in terms of a function $\psi$. Once we do this, wave mechanics is implicit and the framework of special relativity, curvature, symmetry and discreteness is enough to determine the details. Our quantum mechanical assumption is therefore that particle dynamics can be formulated in terms of  a function on the space whose curvature is the electromagnetic force. The proposal that a classical particle should be represented by a wave packet is then the best that we can do within the resulting theory.

It is a fundamental dynamical idea that force is curvature and one that should be acceptable as a point of departure for formulating quantum mechanics. It is at least logically possible that quantum mechanics might have been found in this way in the first place. In formulating general relativity, Einstein was able to use Riemannian geometry, developed some 50 years previously. If, in the interim, mathematicians had already completed the work that led to the theory of fiber bundles, then the discovery of general relativity might have prompted a treatment of electromagnetism along the lines described here, since the other ingredients were in existence by around 1920\footnote{Noether proved her theorem in 1919. The variational principle for electromagnetism was formulated by Weyl (1918) and Born (1909).}.

In practice, general relativity stimulated mathematicians to develop the theory of fiber bundles, a process that would take another 35 years\cite{O'Raifeartaigh:1997ia}. Wave mechanics was arrived at by a different route that started from another idea of Einstein's, that the electromagnetic field is quantized. This idea was the basis for the de Broglie relations that played a significant part in prompting Schr\"odinger to formulate his equation. The two ideas eventually came together in the gauge theory of force in quantum field theory.

\bibliography{references}

\end{document}